\documentclass[12pt,preprint]{aastex}

\usepackage{natbib,amsmath}

\newcommand{\fgrb}{f_{GRB}(L)}
\newcommand{\fdla}{f_{DLA}(L)}

\newcommand{\lya}{Ly$\alpha$}

\def\h2{H$_2$}
\def\f0{$F_0$}

\newcommand{\cm}[1]{\, {\rm cm^{#1}}}

\newcommand{\mnhi}{N_{\rm HI}}

\def\ltp{\left ( \,}

\def\rtp{\, \right  ) }

\begin{document}

\title{Reconciling the Metallicity Distributions 
of Gamma-ray Burst, Damped Lyman-$\alpha$, and Lyman-break Galaxies at $z\approx3$}
\author{Johan P. U. Fynbo \altaffilmark{1},
J. Xavier Prochaska \altaffilmark{2},
Jesper Sommer-Larsen \altaffilmark{3,1}
Miroslava Dessauges-Zavadsky \altaffilmark{4},
Palle M\o ller \altaffilmark{5}
}
\altaffiltext{1}{Dark Cosmology Centre, Niels Bohr Institute, 
University of Copenhagen; Juliane Maries Vej 30, 2100 Copenhagen O, Denmark; 
jfynbo@dark-cosmology.dk}
\altaffiltext{2}{Department of Astronomy and Astrophysics,
UCO/Lick Observatory;
University of California, 1156 High Street,
Santa Cruz, CA 95064; xavier@ucolick.org}
\altaffiltext{3}{Excellence Cluster Universe, Technische Universit\"at 
M\"unchen; Boltz-manstr. 2, D-85748 Garching, Germany}
\altaffiltext{4}{Observatoire de Gen\`eve, 51 Ch. des Maillettes,
1290 Sauverny, Switzerland}
\altaffiltext{5}{European Southern Observatory, Karl-Scharschild-strasse 2, D-85748 Garching bei M\"unchen}
\begin{abstract}
We test the hypothesis that the host galaxies of long-duration 
gamma-ray bursts (GRBs) as well as quasar-selected
damped \lya\ (DLA) systems are drawn from the population of
UV-selected star-forming, high $z$ galaxies (generally referred to
as Lyman-break galaxies). Specifically, we compare
the metallicity distributions of the GRB and DLA populations
against simple models where these galaxies are 
drawn randomly from the distribution of star-forming galaxies
according to their star-formation rate and HI cross-section respectively.
We find that it is possible to match both observational
distributions assuming very
simple and constrained relations between luminosity, metallicity and
HI sizes. The simple model can be tested by observing the luminosity
distribution of GRB host galaxies and by measuring the luminosity 
and impact parameters of DLA selected galaxies as a function of metallicity.
Our results support the expectation that GRB and DLA samples, in
contrast with magnitude limited surveys, provide an almost
complete census of $z\approx3$ star-forming galaxies that are not heavily 
obscured.
\end{abstract} 

\section{Introduction}

The past 10 years has marked the emergence of extensive observational analysis
of high redshift ($z>2$) galaxies. This remarkable and rapid advance was
inspired by new technologies in space and ground-base facilities for deep
imaging, clever approaches to target selection, and the arrival of 10 m class
ground-based telescopes for spectroscopic confirmation. A plethora of
classes are now surveyed, each named for the observational technique that
selects the galaxies: the \lya\ emitters (Hu et al.\ 1998), the Lyman break
galaxies (LBGs, Steidel et al.\ 2003), sub-mm galaxies (e.g., Chapman et al.\
2005), distant red galaxies (van Dokkum et al.\ 2006), damped \lya\ (DLA)
systems (Wolfe et al.\ 2005), extremely red objects (Cimatti et al.\ 2003),
long-duration $\gamma$-ray burst (GRB) host galaxies (e.g., Fruchter et al.\
2006), \ion{Mg}{2} absorbers, radio galaxies (Miley \& De Breuck 2008), quasar
(QSO) host galaxies, etc. Large, dedicated surveys have identified in some
cases thousands of these galaxies providing a direct view into the processes of
galaxy formation in the young universe.

%[this paragraph may not be necessary]
%Within the framework of modern cosmology, each of these galaxies
%is presumed to reside within a virialized dark matter halo. This
%assertion is motivated by the high surface
%densities required to produce the observed stars and/or gas 
%that imply dark matter overdensities $\delta \rho/\rho$
%far above the cosmic mean,
%i.e.\ $\delta \rho / \rho \gg 200$.  In
%Direct evidence for this paradigm
%has been presented through the clustering analysis
%\citep{adelberger, chapman, cwg+06}.
%where the observed clustering signal indicates overdensities 
%consistent with predictions from cosmological modeling of
%dark matter halos.  Furthermore, the gas dynamics for
%a subset of individual galaxies describe potential wells
%representative of massive halos \citep{genzel}.

Because of the significant differences in the sample selection of high $z$
galaxies, there has been a tendency by observers to treat each population
separately and/or contrast the populations. However, the various
populations will overlap to some extent and it is important to understand how
\citep[see also][]{Adelberger00,moller,fynbo03,reddy05}.
%Indeed, allowing for a finite volume density of dark matter
%halos which can host high $z$ galaxies, there must be significant
%overlap. Assuming the WMAP06 cosmology at $z=3$, there are 22~Mpc$^{-3}$ dark
%matter halos with $M> 10^8 M_\odot$.
%Contrast this value with the volume density of \lya\ emitters of 
%$\sim$2~Mpc$^{-3}$ \cite{rauch} and
%Lyman break galaxies with $R$ magnitude less than 27\,mag of
%0.4~Mpc$^{-3}$ \cite{rss+07} [PLEASE CHECK THESE NUMBERS].
%In this paper, we will explore the relationship between three of
%these populations: the star-forming galaxies traced by UV-emission (LBGs),
%the galaxies identified by ISM absorption toward bright
%background sources (DLA systems) and the GRB host galaxies.

Of the various galactic populations discovered at $z>2$ to date,
only two offer the opportunity to study the interstellar medium
at a precision comparable to the Galaxy and its nearest neighbors:
the damped \lya\ systems intervening quasar sightlines (QSO-DLA)
and the host galaxies of GRBs which exhibit bright afterglows
(GRB-DLA\footnote{A small fraction of these GRB afterglow spectra
have \ion{H}{1} column densities in the host galaxy that are below
the standard DLA definition $\mnhi = 2\times10^{20}\cm{-2}$ 
(Jakobsson et al.\ 2006a; Dessauges-Zavadsky al.\ 2008, in prep).  
These GRB-LLS will not be considered here.
Similarly, the QSO absorption line systems with \ion{H}{1} column densities 
between $10^{17}$ and $2\times10^{20} \cm{-2}$ 
(the empirical threshold for the DLAs) will not be considered here.}).  
These galaxies are characterized by a bright background source
which probes the gas along the sightline to Earth. In the case of QSO-DLA
it is the background QSO while for GRB-DLA it is the afterglow of
the GRB located within the host galaxy. Hence, the GRB-DLA will not
probe the full line-of-sight through the host. The gas
in the ISM imprints signatures of the total \ion{H}{1} column
density, the metal content, the ionization state, the velocity fields,
and the molecular fraction along the line-of-sight. Because the galaxies 
are identified
in absorption, there is no formal magnitude limit for the associated stellar
populations. In this respect, they may trace a large dynamic range
in stellar mass, morphology, star formation rate, etc.
%Because the DLA are likely to trace galaxies fainter
%than magnitude-limited surveys \citep[e.g.][]{fynbo99} they are often 
%viewed as unrepresentative of high $z$ galaxies.  Ironically, 
%the DLA galaxies may be the only population which
%include all of the star-forming systems at $z \sim 3$.

The connection between long-duration GRBs and star-forming galaxies has been
empirically established.  At large redshift, there is an exclusive coincidence
of GRBs with actively star-forming galaxies \citep[e.g.,][]{hogg99,
bloom02,fruchter06}, the majority of which show elevated specific
star-formation rates \citep{chg04}.  At low $z$, there is a direct link
between GRBs and massive stars via the detection of spatially and temporally
coinciding core-collapse supernovae
\citep[][but see also Fynbo et al. 2006]{hsm+03,smg+03}.
 Furthermore, GRBs are in some cases found in Wolf-Rayet
galaxies \citep{hfs+06}, which further strengthens the link between 
GRBs and massive star-formation.
The simplest hypothesis, therefore, is that these galaxies uniformly
sample high $z$ galaxies according to star-formation rate, i.e.\ $\fgrb \propto
{\rm SFR}(L) \phi(L)$, where $\phi(L)$ is the luminosity function.  

The link between QSO-DLA and star-forming galaxies is
less direct, primarily because the bright background quasar precludes the easy
detection of stellar light.  Nevertheless, the presence of heavy metals in all QSO-DLAs
(and dust in the majority) indicates at least prior star-formation
\citep{pgw+03}.
Furthermore, the observation of \ion{C}{2}* absorption suggests heating of the
ISM by far-UV photons from ongoing star-formation in at least half of the
sample \citep{wpg03,wolfe04}.  Finally, a handful of QSO-DLA have been detected
in emission and exhibit properties similar to low luminosity LBGs
\citep{moller}. In contrast to the GRB-DLA, however, the QSO-DLA are selected
according to their covering fraction on the sky, i.e.\ the probability of
detection is the convolution of the \ion{H}{1} cross-section with the
luminosity function: $\fdla \propto \sigma_{HI}(L) \phi(L)$.  While both
populations of DLAs may be drawn from the full sample of star-forming galaxies,
their distribution functions would only be the same if $\sigma_{HI}(L) \propto
SFR(L)$ (see also Chen et al.\ 2000).

In this paper, we will test these ideas by comparing the observed
metallicity distributions of QSO-DLA and GRB-DLA with simple predictions
based on empirical measurements of star-forming galaxies at $z=3$.
Specifically, we combine the luminosity function of UV-selected
galaxies (LBGs), a metallicity/luminosity relation, and a
simple prescription for the radial distribution of \ion{H}{1} gas
with luminosity to predict the metallicity distributions of the
QSO-DLA and GRB-DLA. We then compare these models with current
observation to test this general picture.
This analysis is similar in spirit to the studies of \cite{fynbo99}
who combined the LBG luminosity function with a Holmberg relation 
for $\sigma_{HI}$ to predict the luminosities and impact parameters
of QSO-DLA galaxies, \cite{jakobsson} who compared the luminosity distribution
function of GRB host galaxies with the LBG luminosity function, and Chen et
al.\ (2005) and Zwaan et al.\ (2005) who reconciled the properties of local
galaxies with the QSO-DLA cross-section and metal abundances.

The paper is structured in the following way: In Sect.~2 we describe our
methodology, in Sect.~3 our results and in Sect.~4 a discussion of the
results obtained and the uncertainties behind the analysis.

Throughout the paper we assume a cosmology with
$\Omega_\Lambda = 0.70, \Omega_m = 0.30, h=0.7$.

\section{Methodology and Analysis}

\begin{figure}
\epsscale{1.0}
\includegraphics[width=5in, angle=90]{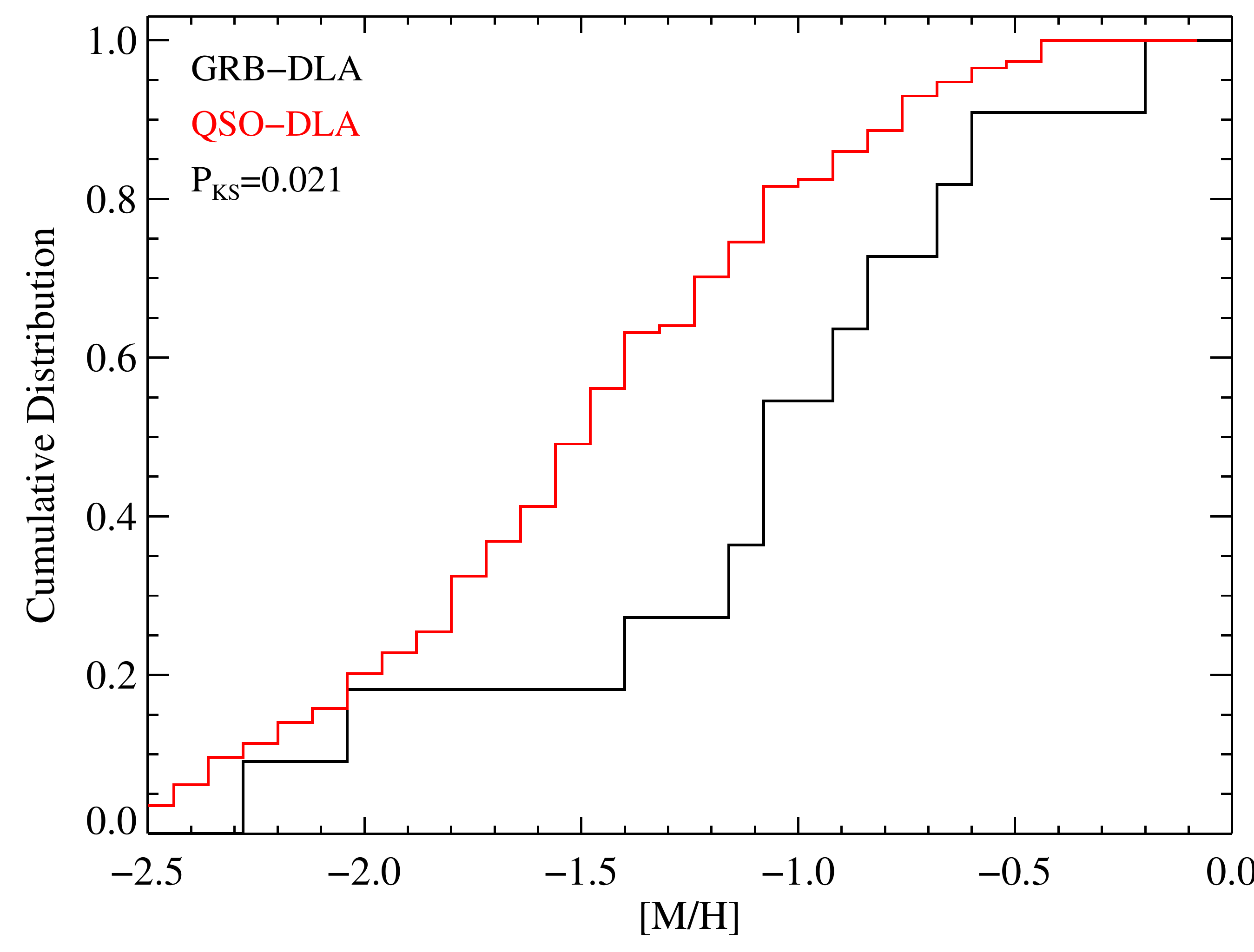}
\caption{The histograms show the cumulative distribution of QSO-DLA and GRB-DLA
metallicities in the statistical samples compiled by \cite{pgw+03} and
\cite{pcd+07}. As seen, the GRB-DLA metallicities are systematically higher
than the QSO-DLA metallicities.
}
\label{fig:cumdata}
\end{figure}

\begin{figure*}
%\plotone{../Figures/hdfn.pdf}
%\epsscale{1.19}
\plotone{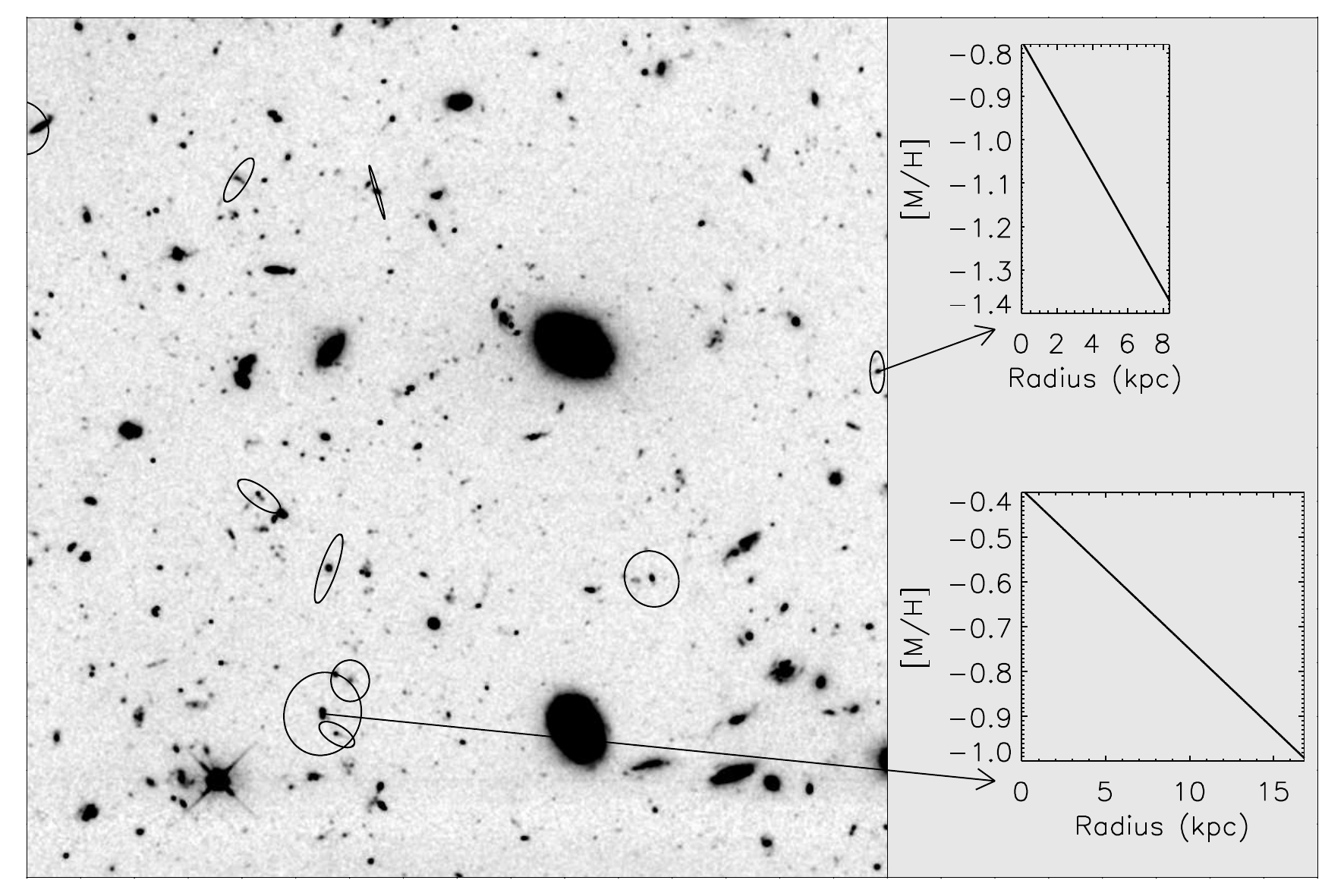}
\caption{A subfield from the HDF North $R$-band image. We have selected
LBGs with redshift between 2.8 and 3.2 from the catalog of photometric
redshifts of Fern\'andez-Soto et al.\ (1999). Over-plotted on each LBGs
is the extent of a randomly inclined HI disk with a radius given in the
text. Note that the majority of the total QSO-DLA absorption cross-section
is caused by fainter galaxies than those shown here. On the right we show 
the radial metallicity profile from the centre to $R_{HI}$ in our model 
for two of the galaxies.
}
\label{fig:illu1}
\end{figure*}

The goal of this paper is to test the hypothesis that
galaxies hosting QSO-DLA and GRB-DLA are drawn from the (same) general
population of star-forming galaxies at high $z$, according to 
\ion{H}{1} cross-section and SFR respectively. We will approach
this hypothesis in a simple, empirical fashion.

Our observational constraint will be the metallicity distribution function of
the two populations shown in Fig.~\ref{fig:cumdata}. The metallicity values are
measured in the same manner for these DLAs: the \ion{H}{1} column density is
derived from a Voigt profile analysis of the \lya\ transition and the metal
column density is estimated from low-ion transitions of non-refractory (or
mildly refractory) elements, e.g.\ S, Si and Zn.  In the following, we will
adopt the statistical samples compiled by \cite{pgw+03} and \cite{pcd+07} for
the QSO-DLA and GRB-DLA respectively.  The redshift range for both samples
is approximately 2--4. We emphasize that nebular measurements of local and
distant galaxies are generally in good agreement with the ISM values
\citep{bowen05,skm+05,pettini02}. For the GRB-DLA, this is a reasonable
expectation because one probes gas surrounding the star-forming region that
hosted the GRB event \citep{pcb06}.  For the QSO-DLA, however, one expects the
ISM probed by these sightlines to be at systematically larger impact parameter,
i.e.\ away from central star-forming regions.  If a radial metallicity gradient
exists in high $z$ galaxies, then the QSO-DLA would give systematically lower
metallicity values than the nebular measurements.  We address this effect in
the following analysis.  Finally, we note that approximately half of the
GRB-DLA metallicity measurements are lower limits because of line saturation.
Also, the GRB-DLA sample is both heterogeneous
(including GRBs from a range of satellites) and biased towards GRBs
with bright optical afterglows. In Sect.~\ref{bias} we further 
discuss the issue of bias.

In Fig.~\ref{fig:illu1} and Fig.~\ref{fig:illu2} we illustrate our model with a
number of schematic figures. Fig.~\ref{fig:illu1} shows a small 
subfield from the Hubble Deep Field (HDF) North (Williams et al.\ 1996). We have here overplotted 
randomly inclined HI disks with sizes as in our model on galaxies with 
photometric redshifts in the range 2.8--3.2. For two of the galaxies,
we also
show the radial metallicity profile as assumed in our model.
The lower panel in Fig.~\ref{fig:illu2} shows the cumulative distribution 
function for observed R-band magnitude for GRB host galaxies and QSO-DLA
galaxies. It is seen that GRB hosts are expected to be brighter than
QSO-DLA galaxies on average. The middle panel shows the impact parameter
distribution for QSO-DLA galaxies in our model. In particular it is seen
that the highest metallicity QSO-DLA will be the brightest and have 
the largest impact parameters. The top panel shows the metallicity distribution
for QSO-DLA (at the QSO-sightline) and GRB host galaxies in our model.

\subsection{Luminosity Function}

Our model is guided by empirical
observations of star-forming galaxies.
In the following,
we assume that the UV-selected sample of $z \sim 3$
galaxies (i.e.\ LBGs) accurately traces the complete sample
of star-forming galaxies \citep[e.g.,][]{Adelberger00}.  
The key caveat is that UV-selected samples are biased against 
extremely dusty galaxies (e.g., Hughes et al.\ 1998; van Dokkum 
et al.\ 2006).
This bias should, however, be a minor effect for galaxies with low
SFR where dust extinction is largely negligible.  
To this extent, we expect the number density of
galaxies with low SFR is well described by the faint end of the
UV luminosity function determined from LBG samples. At very large SFR, however, 
the UV-selected sample may be incomplete and one must at least
consider contributions from populations that better trace dust
enshrouded, star-forming galaxies (e.g.\ sub-mm sources, extremely
red objects).
We will argue below, however, that the dust bias of
UV-selected samples has a minor effect on our conclusions.

We will adopt the UV luminosity function
measured by \cite{rss+08} which provides the largest compilation
of spectroscopically confirmed $z\sim 3$ LBGs to date. The 
luminosity function for these galaxies measured
at rest-frame 1700\AA\ is well
approximated by a Schechter function, 
\begin{equation}
\phi(L_{1700}) = \phi_* (L_{1700}/L_*)^{\alpha}
\exp(-L_{1700}/L*) 
\label{eqn:lumf}
\end{equation}
with a value of $\alpha$ in the range $-1.6$ to $-2.0$
and $M_* = -20.84\pm0.12$.
%The normalization $\phi_*$ is unimportant to our analysis because
%we will only test the shape of the metallicity distributions.
Regarding our analysis, 
the most important characteristic of this luminosity
function is its very steep faint-end slope.  Even
with $\alpha = -1.6$, galaxies with $L > L_*$
contribute only $\approx 10\%$ of the total UV light.
It is not surprising, therefore,
that most magnitude-limited surveys of high $z$ galaxies trace only
the `tip of the iceberg' of the full galactic population
(see also the discussion in Sommer-Larsen \& Fynbo 2008).
%By number and by intensity, the faint-end dominates the galactic
%population at high $z$.
Note that while the bright end has an exponential cutoff,
there is no empirical constraint that sets
the faint-end limit, $L_{min}$.
This is even true in the local universe where one can attain
very sensitive limits (Jerjen et al.\ 2004).   
\cite{baldry05}, for example, have measured the 
local u-band luminosity function and find no break
in the shape of the faint-end slope down to their 
detection limit of $0.016 L_*$. 
Similarly, \cite{blanton05} examined 
extremely faint galaxies in the Sloan Digital Sky Survey (SDSS)
and identified an upturn in the faint-end slope 
towards $\alpha=-1.5$ at $M>-18$. Hence, there is no evidence
from the local Universe that the faint end turns over. On the contrary,
it seems to have an increasing slope with the faintest 
star-forming dwarfs having luminosities around 0.0001 $L_*$ (Mateo 1998).

At redshifts $z\approx3$, GRB hosts are among the faintest known galaxies. The
most striking case is the $z=3.20$ (Hjorth et al.\ 2003) GRB\,020124 for which
Berger et al.\ (2002) placed a limit of $\mathrm{R=29.5}$ on the magnitude of
the host galaxy. This corresponds to $L<0.01 L_*$. Recently, an indirect probe
of the $z=3$ luminosity function has been carried out via very faint Ly$\alpha$
emitters (Rauch et al.\ 2007). This study finds that a Schechter function with
$\alpha=-1.7$ provides a good fit to about 0.025 $L_*$. There is some
indication of a flattening of the slope between 0.025 and 0.01 $L_*$.  In the
following, we will consider values ranging from $L_{min} = 0.1$ to $10^{-5}
L_*$.

\subsection{Metallicity-Luminosity Relation}

To convert the luminosity function into a metallicity distribution for
comparison with the GRB and DLA populations, we must adopt a
metallicity/luminosity relation. In the local universe, metallicity traces
luminosity in low mass (dwarf) galaxies and `saturates' in brighter galaxies
\citep{dw03,tremonti}.  Similar trends have been found for smaller samples of
intermediate and high $z$ galaxies \citep{kk04,moller04,savaglio05,erb}. Guided
by these empirical trends, we will assume a metallicity/luminosity relation of
the form
\begin{equation}
Z = Z_* \ltp \frac{L}{L_*} \rtp^\beta
\label{eqn:ml}
\end{equation}
This relation has two parameters: a normalization $Z_*$ and a slope $\beta$.
In the following we will assume $Z_* = Z_\odot/2$ which is supported
by the handful of precision LBG metallicity measurements to 
date \citep[e.g.][]{pss+01}.
We adopt $\beta = 0.2$ for the slope which lies central
to the various estimates from local and high $z$ analyses
(Tremonti et al.\ 2004; Erb et al.\ 2006; Ledoux et al.\ 2006; 
Prochaska et al.\ 2007). It is also supported by numerical
simulation of $z=3$ galaxies (Fig.~\ref{sommerfig}).
Although our relation does not level off at $L>L_*$, 
our conclusions are insensitive to super-$L_*$ galaxies.

\begin{figure}
%\epsscale{1.0}
\includegraphics[width=5in]{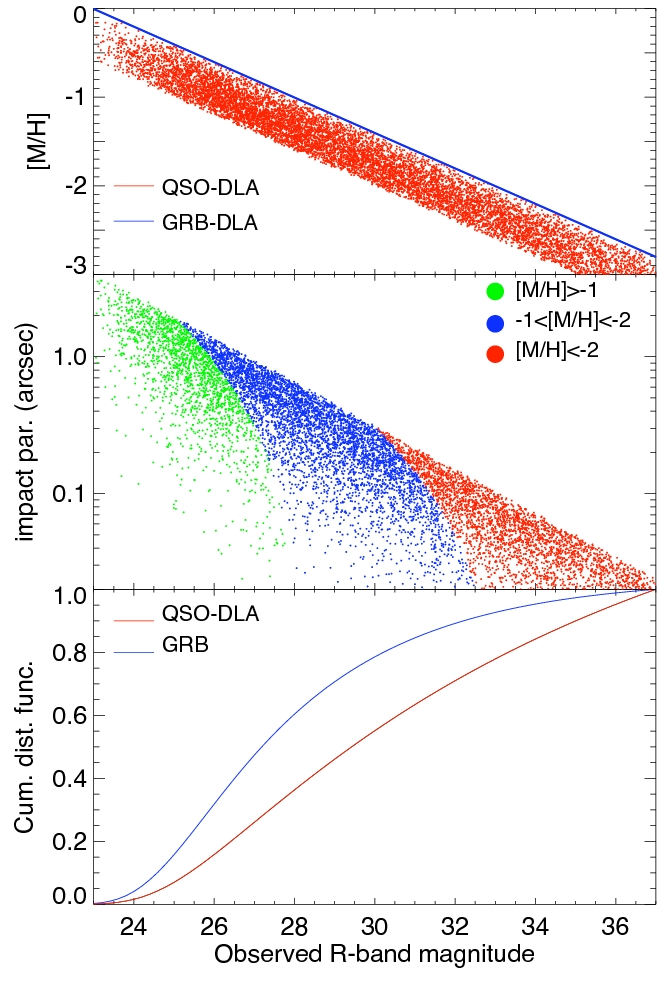}
\caption{
Simulated distributions of luminosity, impact parameter and metallicity 
(from bottom to top) for QSO-DLA and GRB galaxies at $z=3$ in our model.
Here we have used $\alpha=-1.7, \beta=0.2, L_{min}=10^{-4} L^*, t=0.4,$
and $\gamma^*=-0.03$. In the top panel, QSO-DLA have lower metallicities 
than GRB hosts at a given R-band magnitude due to the metallicity gradients.
In the middle panel impact parameters are lower for fainter QSO-DLA galaxies 
due to the Holmberg relation and low metallicity QSO-DLA have lower impact
parameters due to the luminosity-metallicity relation and the Holmberg 
relation. In the lower panel QSO-DLA galaxies are fainter than GRB hosts
as for $t=0.4$ the selection function for QSO-DLA weights fainter galaxies
more than the selection function for GRBs.
}
\label{fig:illu2}
\end{figure}

\begin{figure}
%\epsscale{1.0}
\includegraphics[width=5in]{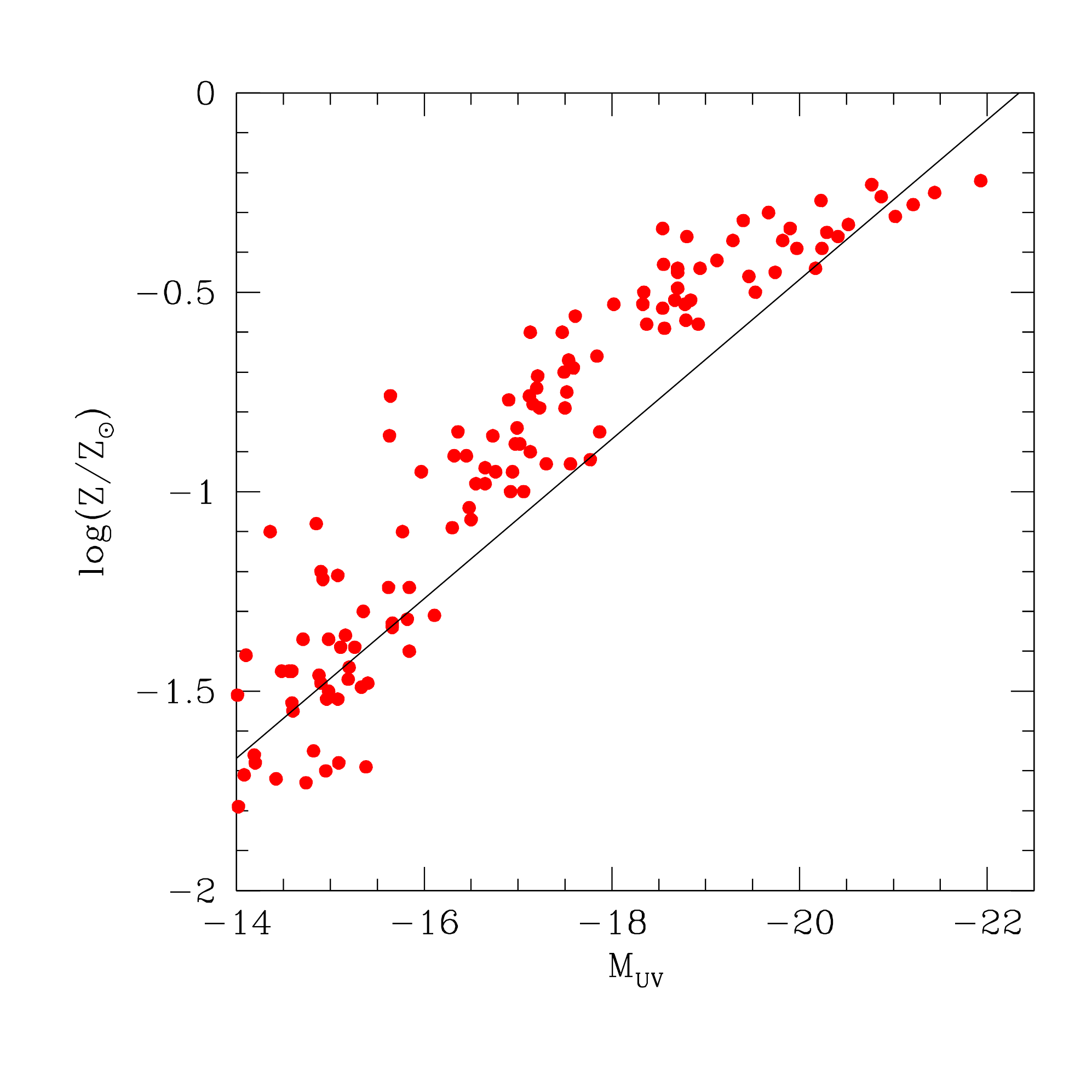}
\caption{The UV luminosity-metallicity relation from the numerical simulations 
of $z=3$ galaxies (from the simulations discussed in Sommer-Larsen \& Fynbo
2008). A slope of $\beta = 0.2$ matches the distribution for simulated galaxies
reasonably well.
}
\label{sommerfig}
\end{figure}

\subsection{Modeling the \ion{H}{1} and Metal Distributions}

To test our hypothesis with QSO-DLA, we must introduce a relation between
\ion{H}{1} cross-section and luminosity, $\sigma_{HI}(L)$. Here we adopt a
Holmberg relation of the form 
\begin{equation}
R_{HI}/R_* = (L/L_*)^t
\label{eqn:holm}
\end{equation}
where $t=0.4$ locally \cite[see, e.g.,][and discussion therein]{wolfe05}.  
For a direct measurement of $t$ for QSO-DLA see also Chen \& Lanzetta (2003).
The value of $R_*$ is fixed so as to reproduce the observed line density of
DLAs $dn/dz(z=3)=0.25$ \citep{phw05}. For randomly inclined disks the relation
between $dn/dz$ and
$R_*$ can be derived for the assumed cosmology following \cite{wolfe86}:
\begin{equation}
\frac{dn}{dz} = (1+z)^3 \frac{dr_{prop}}{dz} \int_{L_{min}}^{\infty}\phi(L)\pi R_{HI}^2/2 dL
\end{equation}
where $\phi$ is the luminosity function and $r_{prop}$ is the proper radial coordinate. Hence, $R_*$ is given by
%\begin{equation}
%R_* = \sqrt{ \frac{dn}{dz} \frac{2H\:\sqrt{(1+z)^2(\Omega_m z+1)-\Omega_{\Lambda}z(z+2)}}{\pi c (1+z)^2 \left[\int_{L_{min}}^{\infty} \phi_* (L/L_{1500})^{1+\alpha+2t} 
%\exp(-L/L_{1500}) \, dL\right]}}
%\end{equation}
\begin{equation}
R_* = \sqrt{ \frac{dn}{dz} \frac{2H_0\:\sqrt{(1+z)^2(\Omega_m z+1)-\Omega_{\Lambda}z(z+2)}}{\pi c (1+z)^2 I }}
\end{equation}
where $I = \int_{L_{min}}^{\infty} \phi_* (L/L_*)^{\alpha+2t} 
\exp(-L/L_*) \, dL$ and $H_0$ is the Hubble parameter.
%For an infinitely thin disc of radius R and with inclination $\theta$
%($\theta=0$ for an edge-on disc) the average impact parameter $b(\theta)$
%can be calculated as the mean radius of an ellipse with major axis $R$ and
%minor
%axis $\sin(\theta)R$. For randomly inclined discs of radius $R$ the
%average impact parameter $\bar{b}$ is then the average value of $b(\theta)$
%weighted with $\sin(\theta)$ from the area and $\cos(\theta)$ from the
%random inclination, i.e.
%\begin{eqnarray}
%\bar{b} & = & 2 R \int_0^{\pi /2}d\theta \sin(\theta)\cos(\theta) b(\theta)
%\nonumber\\
%& = & {} 2 R \int_0^{\pi /2}d\theta \sin(\theta)\cos(\theta)\times {}
%\nonumber\\
%& & {} \left( 4 \int_0^1 dx
%\int_0^{\sin(\theta) \sqrt{1-x^2}}dy \frac{\sqrt{x^2+y^2}}{\pi
%\sin(\theta)}\right)
%\nonumber\\
%& = & {} 0.566 R
%\end{eqnarray}

While the GRB-DLA yield
metallicity measurements of the gas near star-forming regions,
the cross-section to QSO-DLA is maximal at large impact parameters
and these sightlines will show systematically lower metallicity than
the inner SF regions if a metallicity gradient is in place at high $z$.
We will allow for this effect in our model by assuming a metallicity gradient 
\begin{equation}
\gamma = d\log{Z}/dR
\label{eqn:dzdr}
\end{equation}
In the local group values for $\gamma$ measured
for the Galaxy and M33 are $-0.07$ and $-0.05$ dex kpc$^{-1}$ 
\citep{smart97,magrini07}. Lower absolute 
values for $\gamma$ are found for other galaxies, e.g. $\gamma = -0.02$
dex kpc$^{-1}$ for M51 \citep{bresolin04}.
In our model we will follow the work Boissier \& Prantzos (2001) and let 
$\gamma$ depend on luminosity according to the relation:
\begin{equation}
\gamma = \gamma_* R_* / R_{HI}
\label{eqn:dzdr2}
\end{equation}
We will consider $\gamma_*$ values 
ranging from $-0.01$ to $-0.07$\,dex kpc$^{-1}$.

\section{Results}

\begin{figure*}
\epsscale{1.0}
%\plotone{../Figures/fig_grbmonte.pdf}
\includegraphics[width=5in, angle=90]{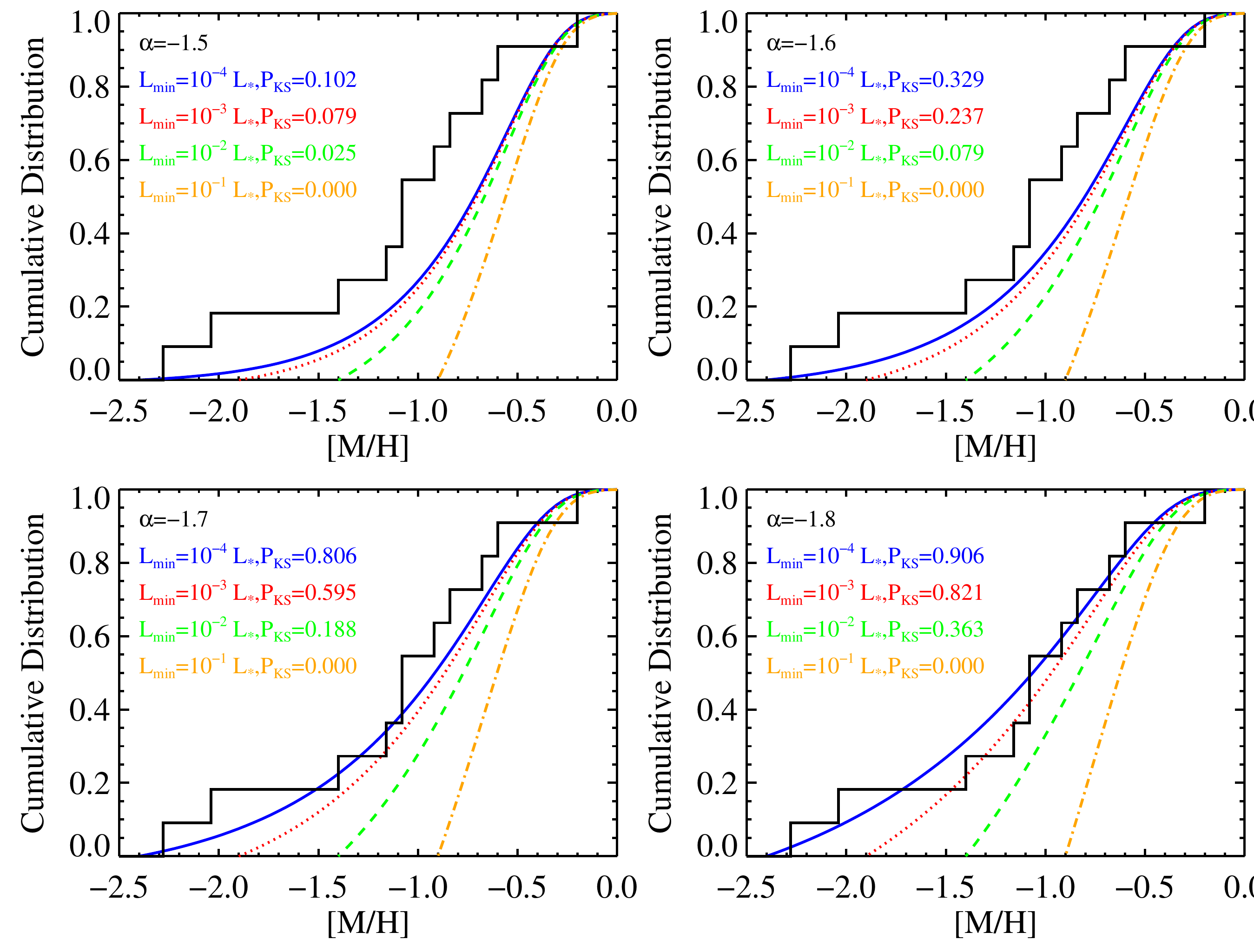}
\caption{The black histograms compare the observed cumulative distribution of GRB-DLA
metallicities against predictions from a series of simple models where one
assumes that GRB host galaxies are drawn according to current SFR 
from the complete set of star-forming galaxies. 
The four panels show 
models assuming faint-end slopes for the luminosity function ranging
from $\alpha=-1.5$ to $\alpha=-1.8$. For each assumed value of $\alpha$ we plot
the model predictions for an assumed lower-limit to the luminosity of
star-forming galaxies $L_{min}$ ranging from $10^{-1} L_*$ (orange dash-dot line) 
to $10^{-4} L_*$ (blue solid line) in steps of factors of 10. 
For each model we derive the probability of a
KS test comparing the model to the data and find good agreement for
the majority of parameter space explored.
%The best model has $\alpha=-1.7$ and
%$L_{min}=0.0001$ $L_*$. 
Models with $\alpha > -1.6$ and/or $L_{min} > 0.01 L_*$
are disfavored by the data.
}
\label{fig:cumgrb}
\end{figure*}

In Fig.~\ref{fig:cumdata} we plot the cumulative metallicity distribution
functions for QSO-DLA and GRB-DLA in the statistical samples compiled by
\cite{pgw+03} and \cite{pcd+07}.  For this figure and the following analyses,
we have incremented the six GRB-DLA lower limits by +0.2\,dex.  This is a
conservative correction; the true value could be significantly 
higher \citep{pro06}.  It is
evident that the GRB-DLA metallicities are higher than the QSO-DLA 
metallicities (see also Savaglio 2006, Fynbo et al.\ 2006b).
In our model, this offset arises naturally from two effects: {\it i)} the
generally higher impact parameters of QSO-DLA relative to GRB-DLA, and {\it
ii)} the fact that QSO-DLA are cross-section selected whereas GRB-DLA are
selected from SFR per luminosity bin. 
As long as the Holmberg parameter $t<0.5$ (eqn.~\ref{eqn:holm})
this will make GRB selected galaxies brighter and
hence more metal rich than QSO-DLA selected galaxies on average.

We can now test the hypothesis that GRB-DLA are uniformly
drawn from the luminosity function of star-forming galaxies
weighted by current SFR, i.e.\ $\fgrb \propto {\rm SFR}(L) \phi(L)$. 
To weight by SFR, we 
assume that SFR(L)~$\propto L_{1700}$,
i.e.\ the far-UV luminosity is an excellent proxy for the SFR
(see also Kennicutt 1998).
Using equation~\ref{eqn:ml} to 
convert the SFR-weighted luminosity function to metallicity,
we derive the cumulative distributions shown in Fig.~\ref{fig:cumgrb}.
%Also present in the figure is the observed cumulative
%distribution of GRB-DLA. As noted above,
%six out of the GRB-DLA have lower limit values to their metallicities.
%In this figure, we have replaced the lower limits with values
%incremented by 0.3\,dex which is a conservative yet reasonable
%estimate of the correction for saturation \citep{pro06,pcd+07}.
It is evident that many of the models are in agreement with the observed
distribution.  We present the $P_{KS}$ values
from a one-sided Kolmogorov-Smirnov test which represent the
probability that the null hypothesis of the observational sample having
been drawn from the model is ruled out.
%A more formal treatment of a distribution with lower limits
%using survival statistics \citep{fn85}
%lead to similar results.
We conclude that the metallicity distribution of GRB-DLA at
$z \sim 3$ is consistent with the prediction 
that GRBs directly trace the star-formation rate and 
that the host metallicities follow a luminosity-metallicity relation.
We also conclude that 
models with $\alpha > -1.6$ and/or $L_{min} > 0.01 L_*$
are disfavored by the data.

%In order to produce predicted metallicity distributions based on our 
%simple model we do the following. For the GRB-DLA distribution
%we draw galaxies randomly using $L\times\phi(L)$ as weight. For each
%galaxy drawn in this way we associate a metallicity using eq.~\ref{eqn:ml}.
%In Fig.~\ref{fig:cumgrb} we show model fits to the GRB-DLA data. As seen,
%we can reproduce the observed distribution with the vales $\alpha=-1.7$
%and $L_{min}=0.0001L_*$. 

For the QSO-DLA,  we draw galaxies randomly using
$\sigma_{HI}(L)$  as the weighting factor, 
i.e.\ $\fdla \propto \sigma_{HI}(L) \phi(L)$ . 
We randomly associate an impact
parameter to the galaxy using again cross-section as weight.  Finally we
associate a metallicity to the sightline using eq.~\ref{eqn:ml} and
eq.~\ref{eqn:dzdr}. In Fig.~\ref{fig:cumqso} we plot models of the
QSO-DLA metallicity distribution. Here we assume $\alpha=-1.7$ and
$L_{min}=0.0001L_*$ for all fits and vary the values of $t$ and $\gamma_*$. 
The best model, which is consistent with the data, has $t=0.4$ and
$\gamma_*=-0.03$ dex kpc$^{-1}$.  We provide the results
for a more exhaustive parameter search in Table~\ref{tab:qsodla}.

%% TABLE
%\input{../Tables/tab_qsodla.tex}
 
\begin{deluxetable}{lcccc}
\tablewidth{0pc}
\tablecaption{RESULTS FOR QSO-DLA. Full table only to be included in the electronic version.\label{tab:qsodla}}
\tabletypesize{\footnotesize}
\tablehead{\colhead{$t$} & \colhead{$\alpha$} &
\colhead{$L_{min}$} & \colhead{$\gamma_*$} & \colhead{P$_{KS}^a$} \\
& & ($L_*$) & (dex kpc$^{-1}$) }
\startdata
0.2&$-1.6$&$10^{-0}$&$-0.01$&$<0.001$\\
0.2&$-1.6$&$10^{-0}$&$-0.02$&$<0.001$\\
0.2&$-1.6$&$10^{-0}$&$-0.03$&$<0.001$\\
0.2&$-1.6$&$10^{-0}$&$-0.04$&$<0.001$\\
0.2&$-1.6$&$10^{-0}$&$-0.05$&0.003\\
0.2&$-1.6$&$10^{-0}$&$-0.06$&0.062\\
0.2&$-1.6$&$10^{-0}$&$-0.07$&0.002\\
0.2&$-1.6$&$10^{-1}$&$-0.01$&$<0.001$\\
0.2&$-1.6$&$10^{-1}$&$-0.02$&$<0.001$\\
0.2&$-1.6$&$10^{-1}$&$-0.03$&0.008\\
0.2&$-1.6$&$10^{-1}$&$-0.04$&0.083\\
0.2&$-1.6$&$10^{-1}$&$-0.05$&0.007\\
0.2&$-1.6$&$10^{-1}$&$-0.06$&$<0.001$\\
0.2&$-1.6$&$10^{-1}$&$-0.07$&$<0.001$\\
0.2&$-1.6$&$10^{-2}$&$-0.01$&$<0.001$\\
0.2&$-1.6$&$10^{-2}$&$-0.02$&$<0.001$\\
0.2&$-1.6$&$10^{-2}$&$-0.03$&$<0.001$\\
0.2&$-1.6$&$10^{-2}$&$-0.04$&$<0.001$\\
0.2&$-1.6$&$10^{-2}$&$-0.05$&$<0.001$\\
0.2&$-1.6$&$10^{-2}$&$-0.06$&$<0.001$\\
0.2&$-1.6$&$10^{-2}$&$-0.07$&$<0.001$\\
0.2&$-1.7$&$10^{-0}$&$-0.01$&$<0.001$\\
0.2&$-1.7$&$10^{-0}$&$-0.02$&$<0.001$\\
0.2&$-1.7$&$10^{-0}$&$-0.03$&$<0.001$\\
0.2&$-1.7$&$10^{-0}$&$-0.04$&$<0.001$\\
0.2&$-1.7$&$10^{-0}$&$-0.05$&$<0.001$\\
0.2&$-1.7$&$10^{-0}$&$-0.06$&0.020\\
0.2&$-1.7$&$10^{-0}$&$-0.07$&0.010\\
0.2&$-1.7$&$10^{-1}$&$-0.01$&$<0.001$\\
0.2&$-1.7$&$10^{-1}$&$-0.02$&$<0.001$\\
0.2&$-1.7$&$10^{-1}$&$-0.03$&0.001\\
0.2&$-1.7$&$10^{-1}$&$-0.04$&0.012\\
0.2&$-1.7$&$10^{-1}$&$-0.05$&0.003\\
0.2&$-1.7$&$10^{-1}$&$-0.06$&$<0.001$\\
0.2&$-1.7$&$10^{-1}$&$-0.07$&$<0.001$\\
0.2&$-1.7$&$10^{-2}$&$-0.01$&$<0.001$\\
0.2&$-1.7$&$10^{-2}$&$-0.02$&$<0.001$\\
0.2&$-1.7$&$10^{-2}$&$-0.03$&$<0.001$\\
0.2&$-1.7$&$10^{-2}$&$-0.04$&$<0.001$\\
0.2&$-1.7$&$10^{-2}$&$-0.05$&$<0.001$\\
0.2&$-1.7$&$10^{-2}$&$-0.06$&$<0.001$\\
0.2&$-1.7$&$10^{-2}$&$-0.07$&$<0.001$\\
0.2&$-1.8$&$10^{-0}$&$-0.01$&$<0.001$\\
0.2&$-1.8$&$10^{-0}$&$-0.02$&$<0.001$\\
0.2&$-1.8$&$10^{-0}$&$-0.03$&$<0.001$\\
0.2&$-1.8$&$10^{-0}$&$-0.04$&$<0.001$\\
0.2&$-1.8$&$10^{-0}$&$-0.05$&$<0.001$\\
0.2&$-1.8$&$10^{-0}$&$-0.06$&$<0.001$\\
0.2&$-1.8$&$10^{-0}$&$-0.07$&0.023\\
0.2&$-1.8$&$10^{-1}$&$-0.01$&$<0.001$\\
0.2&$-1.8$&$10^{-1}$&$-0.02$&$<0.001$\\
0.2&$-1.8$&$10^{-1}$&$-0.03$&$<0.001$\\
0.2&$-1.8$&$10^{-1}$&$-0.04$&0.003\\
0.2&$-1.8$&$10^{-1}$&$-0.05$&$<0.001$\\
0.2&$-1.8$&$10^{-1}$&$-0.06$&$<0.001$\\
0.2&$-1.8$&$10^{-1}$&$-0.07$&$<0.001$\\
0.2&$-1.8$&$10^{-2}$&$-0.01$&$<0.001$\\
0.2&$-1.8$&$10^{-2}$&$-0.02$&$<0.001$\\
0.2&$-1.8$&$10^{-2}$&$-0.03$&$<0.001$\\
0.2&$-1.8$&$10^{-2}$&$-0.04$&$<0.001$\\
0.2&$-1.8$&$10^{-2}$&$-0.05$&$<0.001$\\
0.2&$-1.8$&$10^{-2}$&$-0.06$&$<0.001$\\
0.2&$-1.8$&$10^{-2}$&$-0.07$&$<0.001$\\
0.3&$-1.6$&$10^{-0}$&$-0.01$&$<0.001$\\
0.3&$-1.6$&$10^{-0}$&$-0.02$&$<0.001$\\
0.3&$-1.6$&$10^{-0}$&$-0.03$&$<0.001$\\
0.3&$-1.6$&$10^{-0}$&$-0.04$&0.002\\
0.3&$-1.6$&$10^{-0}$&$-0.05$&0.231\\
0.3&$-1.6$&$10^{-0}$&$-0.06$&0.009\\
0.3&$-1.6$&$10^{-0}$&$-0.07$&$<0.001$\\
0.3&$-1.6$&$10^{-1}$&$-0.01$&$<0.001$\\
0.3&$-1.6$&$10^{-1}$&$-0.02$&$<0.001$\\
0.3&$-1.6$&$10^{-1}$&$-0.03$&0.090\\
0.3&$-1.6$&$10^{-1}$&$-0.04$&0.429\\
0.3&$-1.6$&$10^{-1}$&$-0.05$&0.018\\
0.3&$-1.6$&$10^{-1}$&$-0.06$&$<0.001$\\
0.3&$-1.6$&$10^{-1}$&$-0.07$&$<0.001$\\
0.3&$-1.6$&$10^{-2}$&$-0.01$&0.144\\
0.3&$-1.6$&$10^{-2}$&$-0.02$&0.130\\
0.3&$-1.6$&$10^{-2}$&$-0.03$&0.006\\
0.3&$-1.6$&$10^{-2}$&$-0.04$&$<0.001$\\
0.3&$-1.6$&$10^{-2}$&$-0.05$&$<0.001$\\
0.3&$-1.6$&$10^{-2}$&$-0.06$&$<0.001$\\
0.3&$-1.6$&$10^{-2}$&$-0.07$&$<0.001$\\
0.3&$-1.7$&$10^{-0}$&$-0.01$&$<0.001$\\
0.3&$-1.7$&$10^{-0}$&$-0.02$&$<0.001$\\
0.3&$-1.7$&$10^{-0}$&$-0.03$&$<0.001$\\
0.3&$-1.7$&$10^{-0}$&$-0.04$&$<0.001$\\
0.3&$-1.7$&$10^{-0}$&$-0.05$&0.063\\
0.3&$-1.7$&$10^{-0}$&$-0.06$&0.030\\
0.3&$-1.7$&$10^{-0}$&$-0.07$&$<0.001$\\
0.3&$-1.7$&$10^{-1}$&$-0.01$&$<0.001$\\
0.3&$-1.7$&$10^{-1}$&$-0.02$&$<0.001$\\
0.3&$-1.7$&$10^{-1}$&$-0.03$&0.024\\
0.3&$-1.7$&$10^{-1}$&$-0.04$&0.172\\
0.3&$-1.7$&$10^{-1}$&$-0.05$&0.016\\
0.3&$-1.7$&$10^{-1}$&$-0.06$&$<0.001$\\
0.3&$-1.7$&$10^{-1}$&$-0.07$&$<0.001$\\
0.3&$-1.7$&$10^{-2}$&$-0.01$&0.019\\
0.3&$-1.7$&$10^{-2}$&$-0.02$&$<0.001$\\
0.3&$-1.7$&$10^{-2}$&$-0.03$&$<0.001$\\
0.3&$-1.7$&$10^{-2}$&$-0.04$&$<0.001$\\
0.3&$-1.7$&$10^{-2}$&$-0.05$&$<0.001$\\
0.3&$-1.7$&$10^{-2}$&$-0.06$&$<0.001$\\
0.3&$-1.7$&$10^{-2}$&$-0.07$&$<0.001$\\
0.3&$-1.8$&$10^{-0}$&$-0.01$&$<0.001$\\
0.3&$-1.8$&$10^{-0}$&$-0.02$&$<0.001$\\
0.3&$-1.8$&$10^{-0}$&$-0.03$&$<0.001$\\
0.3&$-1.8$&$10^{-0}$&$-0.04$&$<0.001$\\
0.3&$-1.8$&$10^{-0}$&$-0.05$&0.003\\
0.3&$-1.8$&$10^{-0}$&$-0.06$&0.062\\
0.3&$-1.8$&$10^{-0}$&$-0.07$&0.002\\
0.3&$-1.8$&$10^{-1}$&$-0.01$&$<0.001$\\
0.3&$-1.8$&$10^{-1}$&$-0.02$&$<0.001$\\
0.3&$-1.8$&$10^{-1}$&$-0.03$&0.008\\
0.3&$-1.8$&$10^{-1}$&$-0.04$&0.077\\
0.3&$-1.8$&$10^{-1}$&$-0.05$&0.008\\
0.3&$-1.8$&$10^{-1}$&$-0.06$&$<0.001$\\
0.3&$-1.8$&$10^{-1}$&$-0.07$&$<0.001$\\
0.3&$-1.8$&$10^{-2}$&$-0.01$&$<0.001$\\
0.3&$-1.8$&$10^{-2}$&$-0.02$&$<0.001$\\
0.3&$-1.8$&$10^{-2}$&$-0.03$&$<0.001$\\
0.3&$-1.8$&$10^{-2}$&$-0.04$&$<0.001$\\
0.3&$-1.8$&$10^{-2}$&$-0.05$&$<0.001$\\
0.3&$-1.8$&$10^{-2}$&$-0.06$&$<0.001$\\
0.3&$-1.8$&$10^{-2}$&$-0.07$&$<0.001$\\
0.4&$-1.6$&$10^{-0}$&$-0.01$&$<0.001$\\
0.4&$-1.6$&$10^{-0}$&$-0.02$&$<0.001$\\
0.4&$-1.6$&$10^{-0}$&$-0.03$&$<0.001$\\
0.4&$-1.6$&$10^{-0}$&$-0.04$&0.084\\
0.4&$-1.6$&$10^{-0}$&$-0.05$&0.111\\
0.4&$-1.6$&$10^{-0}$&$-0.06$&$<0.001$\\
0.4&$-1.6$&$10^{-0}$&$-0.07$&$<0.001$\\
0.4&$-1.6$&$10^{-1}$&$-0.01$&$<0.001$\\
0.4&$-1.6$&$10^{-1}$&$-0.02$&$<0.001$\\
0.4&$-1.6$&$10^{-1}$&$-0.03$&0.019\\
0.4&$-1.6$&$10^{-1}$&$-0.04$&0.948\\
0.4&$-1.6$&$10^{-1}$&$-0.05$&0.010\\
0.4&$-1.6$&$10^{-1}$&$-0.06$&$<0.001$\\
0.4&$-1.6$&$10^{-1}$&$-0.07$&$<0.001$\\
0.4&$-1.6$&$10^{-2}$&$-0.01$&$<0.001$\\
0.4&$-1.6$&$10^{-2}$&$-0.02$&$<0.001$\\
0.4&$-1.6$&$10^{-2}$&$-0.03$&0.247\\
0.4&$-1.6$&$10^{-2}$&$-0.04$&0.105\\
0.4&$-1.6$&$10^{-2}$&$-0.05$&$<0.001$\\
0.4&$-1.6$&$10^{-2}$&$-0.06$&$<0.001$\\
0.4&$-1.6$&$10^{-2}$&$-0.07$&$<0.001$\\
0.4&$-1.7$&$10^{-0}$&$-0.01$&$<0.001$\\
0.4&$-1.7$&$10^{-0}$&$-0.02$&$<0.001$\\
0.4&$-1.7$&$10^{-0}$&$-0.03$&$<0.001$\\
0.4&$-1.7$&$10^{-0}$&$-0.04$&0.016\\
0.4&$-1.7$&$10^{-0}$&$-0.05$&0.217\\
0.4&$-1.7$&$10^{-0}$&$-0.06$&$<0.001$\\
0.4&$-1.7$&$10^{-0}$&$-0.07$&$<0.001$\\
0.4&$-1.7$&$10^{-1}$&$-0.01$&$<0.001$\\
0.4&$-1.7$&$10^{-1}$&$-0.02$&$<0.001$\\
0.4&$-1.7$&$10^{-1}$&$-0.03$&0.101\\
0.4&$-1.7$&$10^{-1}$&$-0.04$&0.829\\
0.4&$-1.7$&$10^{-1}$&$-0.05$&0.015\\
0.4&$-1.7$&$10^{-1}$&$-0.06$&$<0.001$\\
0.4&$-1.7$&$10^{-1}$&$-0.07$&$<0.001$\\
0.4&$-1.7$&$10^{-2}$&$-0.01$&$<0.001$\\
0.4&$-1.7$&$10^{-2}$&$-0.02$&0.078\\
0.4&$-1.7$&$10^{-2}$&$-0.03$&0.227\\
0.4&$-1.7$&$10^{-2}$&$-0.04$&0.006\\
0.4&$-1.7$&$10^{-2}$&$-0.05$&$<0.001$\\
0.4&$-1.7$&$10^{-2}$&$-0.06$&$<0.001$\\
0.4&$-1.7$&$10^{-2}$&$-0.07$&$<0.001$\\
0.4&$-1.8$&$10^{-0}$&$-0.01$&$<0.001$\\
0.4&$-1.8$&$10^{-0}$&$-0.02$&$<0.001$\\
0.4&$-1.8$&$10^{-0}$&$-0.03$&$<0.001$\\
0.4&$-1.8$&$10^{-0}$&$-0.04$&0.002\\
0.4&$-1.8$&$10^{-0}$&$-0.05$&0.238\\
0.4&$-1.8$&$10^{-0}$&$-0.06$&0.010\\
0.4&$-1.8$&$10^{-0}$&$-0.07$&$<0.001$\\
0.4&$-1.8$&$10^{-1}$&$-0.01$&$<0.001$\\
0.4&$-1.8$&$10^{-1}$&$-0.02$&$<0.001$\\
0.4&$-1.8$&$10^{-1}$&$-0.03$&0.083\\
0.4&$-1.8$&$10^{-1}$&$-0.04$&0.473\\
0.4&$-1.8$&$10^{-1}$&$-0.05$&0.022\\
0.4&$-1.8$&$10^{-1}$&$-0.06$&$<0.001$\\
0.4&$-1.8$&$10^{-1}$&$-0.07$&$<0.001$\\
0.4&$-1.8$&$10^{-2}$&$-0.01$&0.150\\
0.4&$-1.8$&$10^{-2}$&$-0.02$&0.132\\
0.4&$-1.8$&$10^{-2}$&$-0.03$&0.006\\
0.4&$-1.8$&$10^{-2}$&$-0.04$&$<0.001$\\
0.4&$-1.8$&$10^{-2}$&$-0.05$&$<0.001$\\
0.4&$-1.8$&$10^{-2}$&$-0.06$&$<0.001$\\
0.4&$-1.8$&$10^{-2}$&$-0.07$&$<0.001$\\
0.5&$-1.6$&$10^{-0}$&$-0.01$&$<0.001$\\
0.5&$-1.6$&$10^{-0}$&$-0.02$&$<0.001$\\
0.5&$-1.6$&$10^{-0}$&$-0.03$&$<0.001$\\
0.5&$-1.6$&$10^{-0}$&$-0.04$&0.509\\
0.5&$-1.6$&$10^{-0}$&$-0.05$&0.003\\
0.5&$-1.6$&$10^{-0}$&$-0.06$&$<0.001$\\
0.5&$-1.6$&$10^{-0}$&$-0.07$&$<0.001$\\
0.5&$-1.6$&$10^{-1}$&$-0.01$&$<0.001$\\
0.5&$-1.6$&$10^{-1}$&$-0.02$&$<0.001$\\
0.5&$-1.6$&$10^{-1}$&$-0.03$&$<0.001$\\
0.5&$-1.6$&$10^{-1}$&$-0.04$&0.955\\
0.5&$-1.6$&$10^{-1}$&$-0.05$&$<0.001$\\
0.5&$-1.6$&$10^{-1}$&$-0.06$&$<0.001$\\
0.5&$-1.6$&$10^{-1}$&$-0.07$&$<0.001$\\
0.5&$-1.6$&$10^{-2}$&$-0.01$&$<0.001$\\
0.5&$-1.6$&$10^{-2}$&$-0.02$&$<0.001$\\
0.5&$-1.6$&$10^{-2}$&$-0.03$&0.003\\
0.5&$-1.6$&$10^{-2}$&$-0.04$&0.857\\
0.5&$-1.6$&$10^{-2}$&$-0.05$&$<0.001$\\
0.5&$-1.6$&$10^{-2}$&$-0.06$&$<0.001$\\
0.5&$-1.6$&$10^{-2}$&$-0.07$&$<0.001$\\
0.5&$-1.7$&$10^{-0}$&$-0.01$&$<0.001$\\
0.5&$-1.7$&$10^{-0}$&$-0.02$&$<0.001$\\
0.5&$-1.7$&$10^{-0}$&$-0.03$&$<0.001$\\
0.5&$-1.7$&$10^{-0}$&$-0.04$&0.263\\
0.5&$-1.7$&$10^{-0}$&$-0.05$&0.024\\
0.5&$-1.7$&$10^{-0}$&$-0.06$&$<0.001$\\
0.5&$-1.7$&$10^{-0}$&$-0.07$&$<0.001$\\
0.5&$-1.7$&$10^{-1}$&$-0.01$&$<0.001$\\
0.5&$-1.7$&$10^{-1}$&$-0.02$&$<0.001$\\
0.5&$-1.7$&$10^{-1}$&$-0.03$&0.003\\
0.5&$-1.7$&$10^{-1}$&$-0.04$&0.945\\
0.5&$-1.7$&$10^{-1}$&$-0.05$&0.003\\
0.5&$-1.7$&$10^{-1}$&$-0.06$&$<0.001$\\
0.5&$-1.7$&$10^{-1}$&$-0.07$&$<0.001$\\
0.5&$-1.7$&$10^{-2}$&$-0.01$&$<0.001$\\
0.5&$-1.7$&$10^{-2}$&$-0.02$&$<0.001$\\
0.5&$-1.7$&$10^{-2}$&$-0.03$&0.021\\
0.5&$-1.7$&$10^{-2}$&$-0.04$&0.529\\
0.5&$-1.7$&$10^{-2}$&$-0.05$&$<0.001$\\
0.5&$-1.7$&$10^{-2}$&$-0.06$&$<0.001$\\
0.5&$-1.7$&$10^{-2}$&$-0.07$&$<0.001$\\
0.5&$-1.8$&$10^{-0}$&$-0.01$&$<0.001$\\
0.5&$-1.8$&$10^{-0}$&$-0.02$&$<0.001$\\
0.5&$-1.8$&$10^{-0}$&$-0.03$&$<0.001$\\
0.5&$-1.8$&$10^{-0}$&$-0.04$&0.091\\
0.5&$-1.8$&$10^{-0}$&$-0.05$&0.135\\
0.5&$-1.8$&$10^{-0}$&$-0.06$&$<0.001$\\
0.5&$-1.8$&$10^{-0}$&$-0.07$&$<0.001$\\
0.5&$-1.8$&$10^{-1}$&$-0.01$&$<0.001$\\
0.5&$-1.8$&$10^{-1}$&$-0.02$&$<0.001$\\
0.5&$-1.8$&$10^{-1}$&$-0.03$&0.019\\
0.5&$-1.8$&$10^{-1}$&$-0.04$&0.935\\
0.5&$-1.8$&$10^{-1}$&$-0.05$&0.009\\
0.5&$-1.8$&$10^{-1}$&$-0.06$&$<0.001$\\
0.5&$-1.8$&$10^{-1}$&$-0.07$&$<0.001$\\
0.5&$-1.8$&$10^{-2}$&$-0.01$&$<0.001$\\
0.5&$-1.8$&$10^{-2}$&$-0.02$&$<0.001$\\
0.5&$-1.8$&$10^{-2}$&$-0.03$&0.245\\
0.5&$-1.8$&$10^{-2}$&$-0.04$&0.105\\
0.5&$-1.8$&$10^{-2}$&$-0.05$&$<0.001$\\
0.5&$-1.8$&$10^{-2}$&$-0.06$&$<0.001$\\
0.5&$-1.8$&$10^{-2}$&$-0.07$&$<0.001$\\
0.6&$-1.6$&$10^{-0}$&$-0.01$&$<0.001$\\
0.6&$-1.6$&$10^{-0}$&$-0.02$&$<0.001$\\
0.6&$-1.6$&$10^{-0}$&$-0.03$&$<0.001$\\
0.6&$-1.6$&$10^{-0}$&$-0.04$&0.835\\
0.6&$-1.6$&$10^{-0}$&$-0.05$&$<0.001$\\
0.6&$-1.6$&$10^{-0}$&$-0.06$&$<0.001$\\
0.6&$-1.6$&$10^{-0}$&$-0.07$&$<0.001$\\
0.6&$-1.6$&$10^{-1}$&$-0.01$&$<0.001$\\
0.6&$-1.6$&$10^{-1}$&$-0.02$&$<0.001$\\
0.6&$-1.6$&$10^{-1}$&$-0.03$&$<0.001$\\
0.6&$-1.6$&$10^{-1}$&$-0.04$&0.884\\
0.6&$-1.6$&$10^{-1}$&$-0.05$&$<0.001$\\
0.6&$-1.6$&$10^{-1}$&$-0.06$&$<0.001$\\
0.6&$-1.6$&$10^{-1}$&$-0.07$&$<0.001$\\
0.6&$-1.6$&$10^{-2}$&$-0.01$&$<0.001$\\
0.6&$-1.6$&$10^{-2}$&$-0.02$&$<0.001$\\
0.6&$-1.6$&$10^{-2}$&$-0.03$&$<0.001$\\
0.6&$-1.6$&$10^{-2}$&$-0.04$&0.814\\
0.6&$-1.6$&$10^{-2}$&$-0.05$&$<0.001$\\
0.6&$-1.6$&$10^{-2}$&$-0.06$&$<0.001$\\
0.6&$-1.6$&$10^{-2}$&$-0.07$&$<0.001$\\
0.6&$-1.7$&$10^{-0}$&$-0.01$&$<0.001$\\
0.6&$-1.7$&$10^{-0}$&$-0.02$&$<0.001$\\
0.6&$-1.7$&$10^{-0}$&$-0.03$&$<0.001$\\
0.6&$-1.7$&$10^{-0}$&$-0.04$&0.701\\
0.6&$-1.7$&$10^{-0}$&$-0.05$&$<0.001$\\
0.6&$-1.7$&$10^{-0}$&$-0.06$&$<0.001$\\
0.6&$-1.7$&$10^{-0}$&$-0.07$&$<0.001$\\
0.6&$-1.7$&$10^{-1}$&$-0.01$&$<0.001$\\
0.6&$-1.7$&$10^{-1}$&$-0.02$&$<0.001$\\
0.6&$-1.7$&$10^{-1}$&$-0.03$&$<0.001$\\
0.6&$-1.7$&$10^{-1}$&$-0.04$&0.955\\
0.6&$-1.7$&$10^{-1}$&$-0.05$&$<0.001$\\
0.6&$-1.7$&$10^{-1}$&$-0.06$&$<0.001$\\
0.6&$-1.7$&$10^{-1}$&$-0.07$&$<0.001$\\
0.6&$-1.7$&$10^{-2}$&$-0.01$&$<0.001$\\
0.6&$-1.7$&$10^{-2}$&$-0.02$&$<0.001$\\
0.6&$-1.7$&$10^{-2}$&$-0.03$&$<0.001$\\
0.6&$-1.7$&$10^{-2}$&$-0.04$&0.910\\
0.6&$-1.7$&$10^{-2}$&$-0.05$&$<0.001$\\
0.6&$-1.7$&$10^{-2}$&$-0.06$&$<0.001$\\
0.6&$-1.7$&$10^{-2}$&$-0.07$&$<0.001$\\
0.6&$-1.8$&$10^{-0}$&$-0.01$&$<0.001$\\
0.6&$-1.8$&$10^{-0}$&$-0.02$&$<0.001$\\
0.6&$-1.8$&$10^{-0}$&$-0.03$&$<0.001$\\
0.6&$-1.8$&$10^{-0}$&$-0.04$&0.501\\
0.6&$-1.8$&$10^{-0}$&$-0.05$&0.002\\
0.6&$-1.8$&$10^{-0}$&$-0.06$&$<0.001$\\
0.6&$-1.8$&$10^{-0}$&$-0.07$&$<0.001$\\
0.6&$-1.8$&$10^{-1}$&$-0.01$&$<0.001$\\
0.6&$-1.8$&$10^{-1}$&$-0.02$&$<0.001$\\
0.6&$-1.8$&$10^{-1}$&$-0.03$&$<0.001$\\
0.6&$-1.8$&$10^{-1}$&$-0.04$&0.952\\
0.6&$-1.8$&$10^{-1}$&$-0.05$&$<0.001$\\
0.6&$-1.8$&$10^{-1}$&$-0.06$&$<0.001$\\
0.6&$-1.8$&$10^{-1}$&$-0.07$&$<0.001$\\
0.6&$-1.8$&$10^{-2}$&$-0.01$&$<0.001$\\
0.6&$-1.8$&$10^{-2}$&$-0.02$&$<0.001$\\
0.6&$-1.8$&$10^{-2}$&$-0.03$&0.003\\
0.6&$-1.8$&$10^{-2}$&$-0.04$&0.859\\
0.6&$-1.8$&$10^{-2}$&$-0.05$&$<0.001$\\
0.6&$-1.8$&$10^{-2}$&$-0.06$&$<0.001$\\
0.6&$-1.8$&$10^{-2}$&$-0.07$&$<0.001$\\
\enddata
\tablenotetext{a}{Probability from KS test analysis.}
\end{deluxetable}

\section{Discussion}

Although equations~\ref{eqn:lumf},\ref{eqn:ml},\ref{eqn:holm}, and
\ref{eqn:dzdr2} include a significant number of free parameters, we
emphasize that the GRB-DLA results, in particular, are sensitive to only
a few of these relations.  For the GRB-DLA, the results depend on four
parameters (and the assumption that $L_{1700}$ traces the SFR): 
the faint-end slope $\alpha$, the metallicity $Z_*$ of an $L=L_*$ galaxy,
the power-law exponent $\beta$ of the metallicity/luminosity relation,
and the minimum luminosity $L_{min}$ of a star-forming galaxy relative to $L_*$.
Both $\alpha$ and $Z_*$ are reasonably well constrained by observations
at $z \approx 3$. 
Although $\beta$ is not strongly constrained at $z=3$, there
is evidence that the local value of 0.2 is also valid 
\citep{ledoux06,erb,pcd+07} (see also Fig.~\ref{sommerfig}).
Therefore, the GRB-DLA results are sensitive to only two relatively
unconstrained parameters. Regarding $L_{min}$, assuming $\alpha \le -1.6$
we find good agreement
between observations and model for $L_{min} < 10^{-3} L_*$ and we cannot
rule out $L_{min} \approx 10^{-2} L_*$. 
In terms of the metallicity/luminosity relation, we have not extensively
explored $\beta$ in part because there is degeneracy in
the metallicity predictions between $\alpha$ and $\beta$.

\begin{figure*}
\includegraphics[width=5in, angle=90]{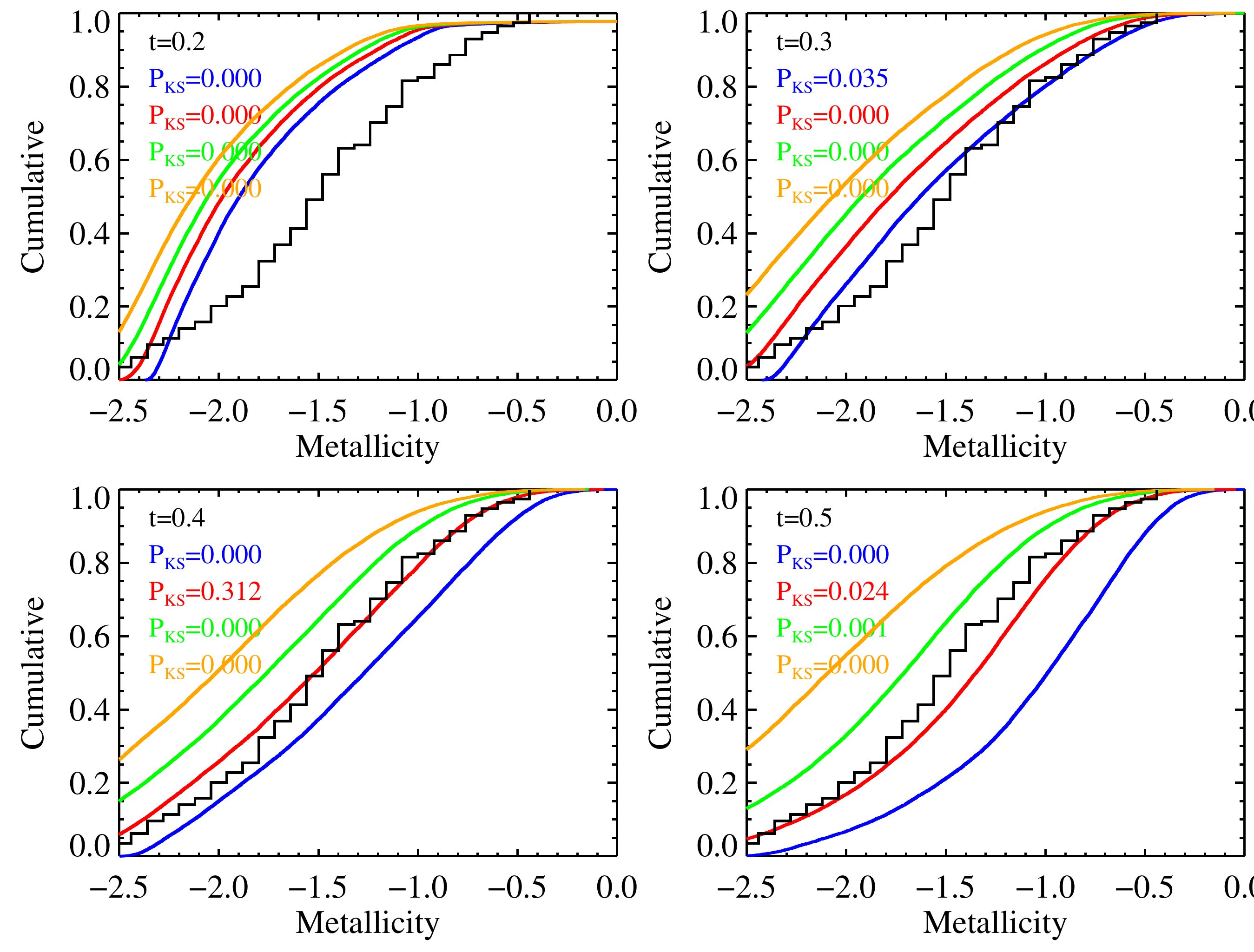}
\caption{The histograms show the cumulative distribution of QSO-DLA
metallicities. The four panels show the predicted metallicity distribution in
our simple model assuming values of $t$ ranging from 0.2 to 0.5.  For each
assumed value of $t$ we plot the model for values of $\gamma_*$ (in units of dex
kpc$^{-1}$) of $-0.07$ (orange), $-0.05$ (green), $-0.03$ (red), and $-0.01$
(blue). For each model we derive the probability of a KS test comparing the
model to the data. Holmberg slopes of $t=0.3$--0.5 and radial abundance
gradients of $\gamma_*=0.01$--0.05 dex kpc$^{-1}$ provide acceptable fits. 
For all models
shown here we have used $\alpha=-1.7$ and  $L_{min}=0.0001$ $L_*$. 
%Models with $L_{min} < 0.0001 L_*$ are disfavored by the data.
}
\label{fig:cumqso}
\end{figure*}

The QSO-DLA models do include additional scaling laws and therefore more free
parameters.  Specifically, we have included a Holmberg relation and metallicity
gradient for the gas in star-forming galaxies.  Neither of these relations have
been constrained observationally at high redshift \citep[however, there is
evidence that abundance gradients exist in some high-$z$ galaxies,
see][]{fschreiber06}; we assumed values from the local universe. 
We also note there is evidence for metallicity gradients in QSO-DLA galaxies
at intermediate redshifts (Chen, Kennicutt \& Rauch 2005).
To this
extent, we do not consider the results shown in Fig.~\ref{fig:cumqso} to be a
special success.  Nevertheless, our results demonstrate that the observations
are reproduced when adopting standard values for these prescriptions from the
local universe.  One implication of this success is that the model may be a
fair representation of average
%the metallicity distributions of QSO-DLA, GRB-DLA and LBGs. The model is
%determined by very few parameters: the 3 parameters $\phi_*$, $L_*$ and
%$\alpha$ defining the luminosity function, the Holmberg slope $t$, the
%metallicity at $L_*$, the slope $/beta$ of the L-Z relation, the metallicity
%gradient $\gamma$, the faint end slope of the luminosity function $\alpha$ and
%the minimum luminosity $L_{min}$.  These parameters are all but $\gamma_*$
%constrained directly by observations of high-$z$ galaxies and the model that
%fits the observations has parameter values fully consistent with the
%observational constraints on these parameters. The implication of this is
%success is that the model is likely to be a fair representation of average
scaling relations that must be at place at $z\approx3$.
We wish to caution, however, that there is large degeneracy
between the various relations applied for the QSO-DLA.

%The model is
%determined by very few parameters: the 3 parameters $\phi_*$, $L_*$ and
%$\alpha$ defining the luminosity function, the Holmberg slope $t$, the
%metallicity at $L_*$, the slope $/beta$ of the L-Z relation, the metallicity
%gradient $\gamma$, the faint end slope of the luminosity function $\alpha$ and
%the minimum luminosity $L_{min}$.  These parameters are all but $\gamma$
%constrained directly by observations of high-$z$ galaxies and the model that
%fits the observations has parameter values fully consistent with the
%observational constraints on these parameters. 

\subsection{How biased are the QSO-DLA and GRB-DLA samples?}
\label{bias}

%[JXP:  I think a discussion of dust bias is warranted, but you might
%consider doing less.  My point is that the comparison between model
%prediction and observation is not being driven at the high metallicity
%end right now, i.e. the place where dust bias would matter.  As such,
%I'd just point out that fact, reference a couple of papers and move on.]  

An obvious question is whether our QSO-DLA and GRB-DLA samples are
representative of the underlying galaxy populations?  It has long been
discussed to which extent QSO-DLA samples are biased against dusty (and
hence likely metal rich and/or large $\log{N}$(\ion{H}{1})) systems.  Studies
of radio selected DLAs (free from dust-bias) have found similar column density
and metallicity distributions as for optically selected DLA samples (Ellison et
al.\ 2001, 2004; Ellison, Hall \& Lira 2005; Akerman et al.\ 2005; Jorgenson et
al.\ 2006) showing that any dust bias will be so small that it will not
fundamentally change the conclusions about cross-section and metallicity
distributions inferred from optically selected surveys.

Concerning GRB-DLA there is no dust bias in the detection of the prompt
emission itself as $\gamma$-rays are unaffected by dust. However, the
requirement of an optical afterglow detection from which the redshift the
\ion{H}{1} column density and metal columns can be measured does potentially
exclude very dusty sightlines.  Furthermore, there could be an intrinsic
(astrophysical) bias against high metallicity in GRB production. In the
collapsar model the limit is estimated to be around 0.3 Z$_{\odot}$ (Hirschi et
al.\ 2005; Woosley \& Heger 2005), but this is very dependent on the as yet
poorly understood properties of winds from massive stars (e.g. clumping, Smith
2007). We also note that Wolf \& Podsiadlowski (2007) exclude a metallicity
cut-off below half the solar value based on statistics of host galaxy
luminositues. If true such an intrinsic bias could preferentially exclude
massive, dust-obscured starbursts, that typically seem to be enriched above
this limit (Swinbank et al. 2004), from the GRB samples. Nevertheless, a few
extremely red and luminous GRB hosts have been found (Levan et al.\ 2006;
Berger et al.\ 2007).  So far there are few examples of GRB sightlines with
very large dust columns (for recent examples see Rol et al.\ 2007; Jaunsen et
al.\ 2008; Tanvir
et al.\ 2008). The near, mid and far-IR properties of GRB host galaxies have
been studied by a number of groups (e.g., Chary et al.\ 2002; Le Floc'h et al.\
2003, 2006; Berger et al.\ 2003; Tanvir et al.\ 2004; Priddey et al. 2006;
Castro Cer{\'o}n et al.\ 2007). A few GRB hosts, all at $z<2$, have tentatively
been detected at sub-mm wavelengths, but their inferred UV/optical (bluer,
Gorosabel et al.\ 2003a,b) and dust properties (higher temperatures,
Micha\l{}owski et al.\ 2008) are different than those of sub-mm selected
galaxies. 

We are currently in the process of building up a $\gamma$-ray/X-ray selected
sample of GRBs with the aim of getting control on the issue of bias (Jakobsson
et al. 2006b; Fynbo et al.\ in preparation). Our sample selection excludes
bursts with high foreground extinction and low sun angle. Hence, the sample
includes bursts for which conditions for optical follow-up are favourable, but
the sample is not biased towards optically bright bursts.  Currently, optical
afterglows are detected for about 80\% of the bursts in our sample and there
is no obvious difference in the X-ray absorption properties of GRBs with and
without detected optical afterglows. Therefore, we consider it unlikely that
the current sample of GRB-DLA is strongly biased against high metallicity and
dusty systems, but this is an issue that still needs to be clarified. In terms
of the analysis presented here, we conclude that dust bias will play a minor
role.

%\subsection{Implications of bound on $L_{min}$}
%Not sure there is much to learn there?

%\subsection{Integral constraints}
%As a sanity check we can calculate the total amount of metals in galaxies 
%in our simple model and in particular ask which parts of the luminosity
%function dominates the metal content. 
%
%[JXP:  Note that integral constraints are somewhat tricky here.  We have
%no new constraint on $\phi_*$ from the GRB data and I would hesitate to
%use the DLA metallicities to place any serious constraint on $\phi_*$.]
%
%[Is our model consistent with the total extragalactic background light (e.g. Bernstein et al.)?
%Is the total amount of metals reasonable? Where are the metals in our simple model?
%Total production of UV ionizing photons. Look at the Bianchi et al.\ paper 
%(2001A\&A...376....1B).
%]

\subsection{How can this simple picture be tested?}

The best fitting model has a value of the Holmberg parameter $t$ of 0.4.  For
$t=0.5$, the probability to select a GRB host or a DLA galaxy counterpart with
luminosity $L$ is identical in our model.  For values of $t$ less than 0.5,
QSO-DLA galaxies counterparts are predicted to be on average fainter than GRB
host galaxies. An indication that this indeed may be the case comes from the
fact that Ly$\alpha$ emission is detected more often in GRB-DLA than
QSO-DLA. However, this may also be a result of the on average smaller impact
parameters for GRB sightlines and the fact the GRB hosts are probably selected
to have a high instantaneous specific star-formation rate.

One way to test this model is
to measure the luminosities of GRB and QSO-DLA selected galaxies and 
compare with the predictions of the model. 
Unfortunately, it has proven extremely difficult to detect QSO-DLA galaxy 
counterparts in emission to the glare of the bright background QSOs. 
However, a clear prediction of the model (see Fig.~3) is that the most 
metal rich QSO-DLA should have the brightest galaxy counterparts and 
the largest impact parameters. It would therefore be interesting to 
target a sample of such DLAs and constrain the statistical properties
of the galaxy counterparts. 

The outlook for a statistically significant sample of GRB host galaxy
luminosities is much better (Hjorth et al.\ 2008, in preparation).
In Jakobsson et al.\ 
(2005) it is described how the model can be tested from a sample of
GRB host galaxy luminosities and upper limits. Given the heterogeneous
nature of the current sample of GRB-DLA with metallicity measurements
it is also important to obtain metallicity measurements for a more 
complete sample of GRB-DLA. Instruments like the X-shooter spectrograph
on the European Southern Observatory Very Large Telescope will make 
this feasible in the near future (Kaper et al.\ 2008).

\section{Conclusions}
We find that with a simple model including the luminosity function for LBGs, a
Holmberg relation for $\sigma(L)$, an L-Z relation and a metallicity gradient
with a slope $\gamma$ dependent on luminosity it is possible to reconcile the
metallicity distributions of QSO-DLA, GRB-DLA and LBGs.  In this model the
faint end of the luminosity function plays a very important role. As seen in
the lower panel in Fig.~\ref{fig:illu2} more than 75\% of star-formation
selected galaxies are fainter than the flux limit for LBGs, R=25.5. For QSO-DLA
galaxies the fraction is even higher.  Hence, in this model the GRB and DLA
samples, in contrast with magnitude limited surveys, provide an almost complete
census of $z\approx3$ star-forming galaxies that are not heavily 
obscured.

\acknowledgements
We acknowledge helpful discussions with E. Ramirez-Ruiz and M. Pettini, Andrew
Pontzen, and H.-W. Chen. We also thank the anonymous referee for a very
constructive and helpful report. J. X.  P. is partially supported by NASA/Swift
grants NNG06GJ07G and NNX07AE94G and an NSF CAREER grant (AST-0548180). This
research was supported by the DFG cluster of excellence ``Origin and structure
of the Universe''. The Dark Cosmology Centre is funded by the DNRF.

{}

\end{document}